\documentclass[aps,prl,twocolumn,superscriptaddress,nofootinbib]{revtex4-2}
\usepackage{graphicx}
\usepackage{epstopdf}
\usepackage{amssymb, amsmath}
\usepackage[usenames]{color}
\usepackage{bm}
\usepackage{ulem}
\usepackage{multirow}
\usepackage{threeparttable}
\usepackage{booktabs}
\usepackage{dcolumn}
\usepackage{float}
\usepackage{hyperref}

\hypersetup{
	colorlinks=true,
	linkcolor=blue,
	filecolor=gray,      
	urlcolor=blue,
	citecolor=blue,
}

\date{\today}

\begin{document}

\title{A Geometric Design Principle for $\mathbb{Z}_2$ Topological Phases in\\
Twisted Triangular-Lattice Bilayers}

\begin{abstract}
Twisted van der Waals bilayers provide a versatile platform for moir\'{e} electronic states, yet a transferable symmetry-based principle for time-reversal-invariant $\mathbb{Z}_2$ moir\'{e} bands has remained largely missing. Here we show that triangular-lattice bilayers with symmetry-related stacking minima provide a geometric route to an emergent honeycomb moir\'{e} lattice. Band-edge states derived from the untwisted $\Gamma$ valley are trapped by the reconstructed stacking landscape, forming A/B moir\'{e} orbitals whose inter-domain coupling generates Dirac crossings. Spin--orbit coupling opens a topological gap, yielding an effective Kane--Mele description and a quantum spin Hall phase characterized by a nontrivial $\mathbb{Z}_2$ invariant. First-principles calculations for Janus BiTeBr confirm the robustness of this phase over a broad twist-angle range and demonstrate an electric-field-driven topological transition. Representative triangular-lattice bilayers further establish this symmetry-based design principle as a broadly applicable route to tunable moir\'{e} quantum spin Hall materials.
\end{abstract}

\author{Jiaheng Li $^{*,\dagger}$}
\affiliation{Beijing National Laboratory for Condensed Matter Physics and Institute of Physics, Chinese Academy of Sciences, Beijing 100190, China}

\author{Jiaxuan Liu $^{*}$}
\affiliation{Beijing National Laboratory for Condensed Matter Physics and Institute of Physics, Chinese Academy of Sciences, Beijing 100190, China}
\affiliation{University of Chinese Academy of Sciences, Beijing 100049, China}

\author{Yan Zhang $^{*}$}
\affiliation{Beijing National Laboratory for Condensed Matter Physics and Institute of Physics, Chinese Academy of Sciences, Beijing 100190, China}
\affiliation{University of Chinese Academy of Sciences, Beijing 100049, China}

\author{Zhong Fang}
\affiliation{Beijing National Laboratory for Condensed Matter Physics and Institute of Physics, Chinese Academy of Sciences, Beijing 100190, China}
\affiliation{University of Chinese Academy of Sciences, Beijing 100049, China}

\author{Hongming Weng $^{\dagger}$}
\affiliation{Beijing National Laboratory for Condensed Matter Physics and Institute of Physics, Chinese Academy of Sciences, Beijing 100190, China}
\affiliation{University of Chinese Academy of Sciences, Beijing 100049, China}
\affiliation{Condensed Matter Physics Data Center of Chinese Academy of Sciences, Beijing 100190, China}

\author{Quansheng Wu $^{\dagger}$}
\affiliation{Beijing National Laboratory for Condensed Matter Physics and Institute of Physics, Chinese Academy of Sciences, Beijing 100190, China}
\affiliation{University of Chinese Academy of Sciences, Beijing 100049, China}

\begingroup
\renewcommand{\thefootnote}{}
\footnotetext{
$^*$ These authors contributed equally to this work.\\
$^\dagger$ To whom correspondence should be addressed; \\
\hspace*{1.2em}E-mail: lijiaheng@iphy.ac.cn, hmweng@iphy.ac.cn,\\
\hspace*{1.2em}and quansheng.wu@iphy.ac.cn.
}
\endgroup

\maketitle

\textit{Introduction.---}Moir\'{e} superlattices have emerged as a versatile platform for engineering electronic states in two-dimensional van der Waals materials \cite{carr2020electronic, andrei2021marvels, kennes2021moire}. Interlayer twisting generates long-wavelength moir\'{e} potentials that strongly reconstruct the electronic structure, producing narrow minibands where quantum geometry and interaction effects become dominant \cite{bistritzer2011moire, cao2018unconventional, zhang2020flat, cao2020tunable}. This setting has enabled a broad range of correlated and topological phases, including superconductivity, Mott insulating states, Wigner crystals, and quantum Hall phenomena at zero magnetic field \cite{wang2020correlated, wu2019topological, zeng2023thermodynamic, cai2023signatures, xu2023observation, zhang2024polarization, wang2024fractional, redekop2024direct, jia2024moire, xia2025superconductivity, xu2025multiple, guo2025superconductivity}, underscoring the rich interplay between topology, symmetry, and interactions in moir\'{e} systems.

Despite these advances, a central challenge remains: moir\'{e} topological phases are often difficult to reproduce and predict from microscopic material information alone. Their topology can be highly sensitive to twist angle, stacking configuration, and lattice relaxation, leading to strong sample-to-sample variation and discrepancies between theory and experiment \cite{tang2021geometric, jia2024moire, zhang2024polarization, zhang2024universal, xu2025multiple, yang2024topological, liu2025symmetry}. In the small-angle regime, the large moir\'{e} unit cells further hinder systematic first-principles searches, restricting progress to case-by-case studies \cite{carr2017twistronics, carr2020electronic}. As a result, time-reversal-invariant moir\'{e} topology, including quantum spin Hall and fractional spin Hall phases, remains much less developed than its Chern-band counterpart \cite{stern2016fractional, kang2024evidence}.

In contrast, the discovery of new quantum materials has historically relied on transferable design principles rather than material-specific searches. In crystalline systems, rules based on symmetry-protected band inversion, electron filling constraints, and structural chemistry have enabled the systematic identification of topological insulators and quantum materials families \cite{bernevig2006quantum, zhang2009topological, galanakis2002origin, bartel2019new}. However, an analogous predictive framework for $\mathbb{Z}_2$ topology in moir\'{e} systems is still missing, leaving time-reversal-invariant moir\'{e} phases without a general geometric design principle.

In this Letter, we uncover a geometry-driven design principle for realizing time-reversal-invariant $\mathbb{Z}_2$ moir\'{e} topology in triangular-lattice bilayers. The key idea is that symmetry-related stacking minima in the untwisted bilayer evolve, under twisting and relaxation, into alternating A and B moir\'{e} domains, which provide the two trapping sites of an emergent honeycomb lattice for $\Gamma$-valley band-edge states. First-principles calculations for BiTeBr demonstrate this mechanism, with $\Gamma$-valley minibands carrying a nontrivial $\mathbb{Z}_2$ invariant over a broad range of twist angles. A four-band Wannier construction establishes an extended Kane--Mele description, while an out-of-plane electric field drives a topological phase transition by tuning spin--orbit competition. Additional triangular-lattice bilayers exhibit the same symmetry-organized honeycomb structure, confirming the generality of the design principle.

\begin{figure}[htbp]
    \includegraphics[width=0.95\linewidth]{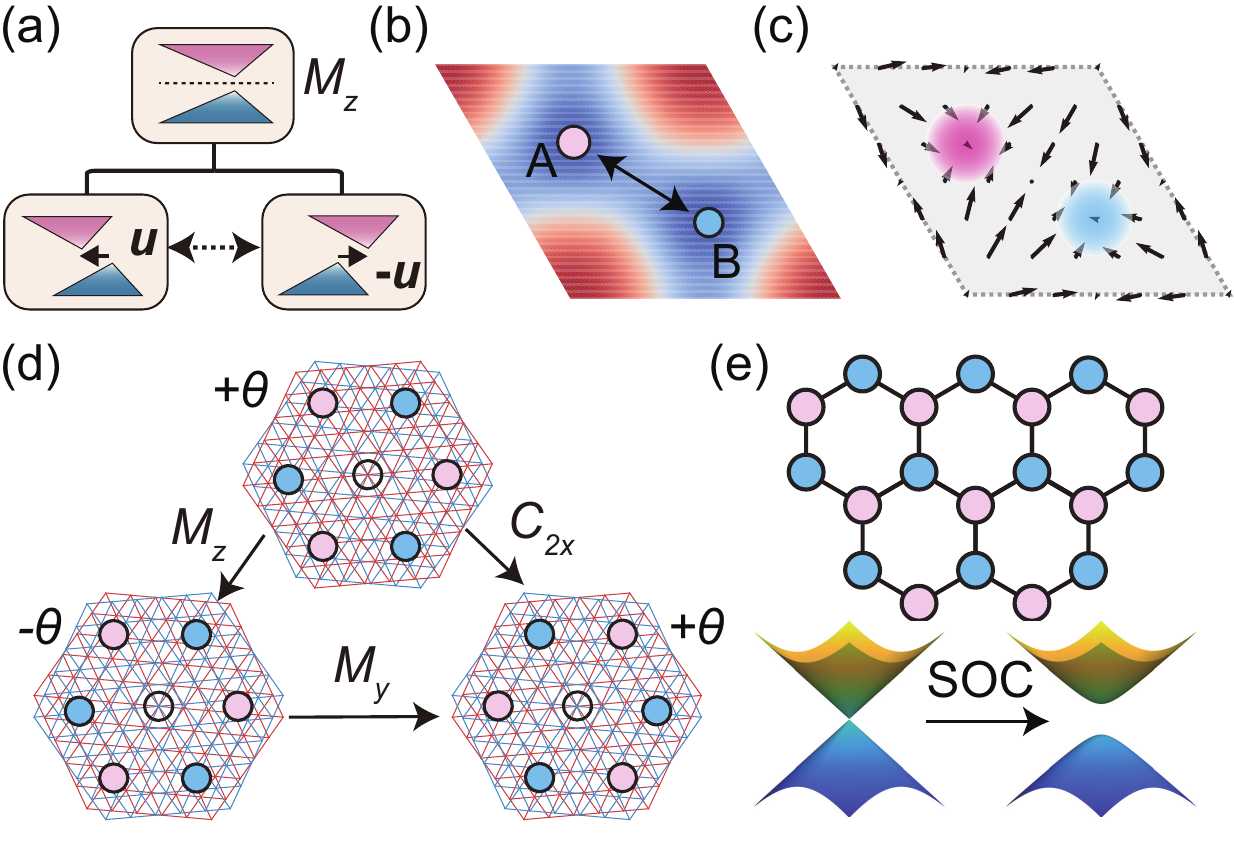}
    \caption{
    Geometric design principle for moir\'e honeycomb reconstruction in triangular-lattice bilayers. 
    (a) In the untwisted bilayer, the layer-exchanging mirror $M_z$ maps a stacking configuration $\mathbf{u}$ to $-\mathbf{u}$, making the two configurations symmetry equivalent. 
    (b) The resulting stacking landscape hosts two symmetry-equivalent degenerate minima A and B.
    (c) Upon twisting, lattice relaxation generates a spatially varying displacement field $\mathbf{u}(\mathbf{r})$. 
    (d) $M_z$ and $M_y$ are broken, while their product $C_{2x}=M_yM_z$ remains as a moir\'{e} symmetry exchanging A and B domains.
    (e) The reconstructed A/B domains form a honeycomb lattice, where inter-sublattice coupling produces Dirac states and spin--orbit coupling opens a gap, yielding a quantum spin Hall phase with nontrivial $\mathbb{Z}_2$.
    }
    \label{fig1_schematic}
\end{figure}

\textit{Geometric design principle.---}
The geometric design principle illustrated in Fig.~\ref{fig1_schematic} originates from symmetry constraints in triangular-lattice bilayers. In the untwisted limit, the layer-exchanging mirror $M_z$ maps a stacking configuration $\mathbf{u}$ to $-\mathbf{u}$, rendering the two configurations symmetry equivalent [Fig.~\ref{fig1_schematic}(a)]. Together with $C_{3z}$, this symmetry selects a pair of equivalent stacking registries, denoted A and B, in the stacking-energy landscape [Fig.~\ref{fig1_schematic}(b)].

Upon twisting, lattice relaxation generates a spatially varying displacement field $\mathbf{u}(\mathbf{r})$ [Fig.~\ref{fig1_schematic}(c)]. In the small-angle regime, the system reconstructs into alternating A/B domains separated by narrow domain walls. The individual mirror symmetries $M_z$ and $M_y$ are generally lost in the moir\'{e} pattern, but their product $C_{2x}=M_yM_z$ remains as a moir\'{e} symmetry that exchanges the two reconstructed domain types [Fig.~\ref{fig1_schematic}(d)]. This residual symmetry is crucial: it keeps the A and B domains energetically equivalent after reconstruction, despite their spatial separation.

As a consequence, band-edge states derived from the untwisted $\Gamma$ valley become confined near the two symmetry-related stacking minima, forming A and B moir\'{e} orbitals on an emergent honeycomb lattice [Fig.~\ref{fig1_schematic}(e)]. Because the $\Gamma$-valley states vary smoothly in real space, they are primarily sensitive to the slowly varying moir\'{e} potential rather than Brillouin-zone folding effects. This makes them a natural low-energy basis for a direct projection onto the moir\'{e} superlattice.

The preserved $C_{2x}=M_yM_z$ symmetry exchanges the two reconstructed domain types and enforces equal onsite energies for the A and B moir\'{e} orbitals. Inter-domain hybridization between neighboring A and B orbitals then generates the honeycomb band connectivity, giving rise to Dirac crossings at the moir\'{e} $K$ points in the spin-independent limit. Spin--orbit coupling further induces symmetry-allowed mass terms that open a time-reversal-preserving topological gap. From the perspective of topological quantum chemistry~\cite{bradlyn2017topological, kruthoff2017topological, song2018quantitative}, this corresponds to a symmetry-driven reorganization of the $\Gamma$-valley band-edge states into a connected honeycomb band representation, realizing a Kane--Mele-type quantum spin Hall phase~\cite{kane2005quantum}. This identifies two microscopic criteria for the design principle: symmetry-related stacking minima that form A/B moir\'{e} domains, and a $\Gamma$-valley band edge whose smooth envelope states can be trapped by the reconstructed stacking landscape.

\begin{figure}[htbp]
\includegraphics[width=0.98\linewidth]{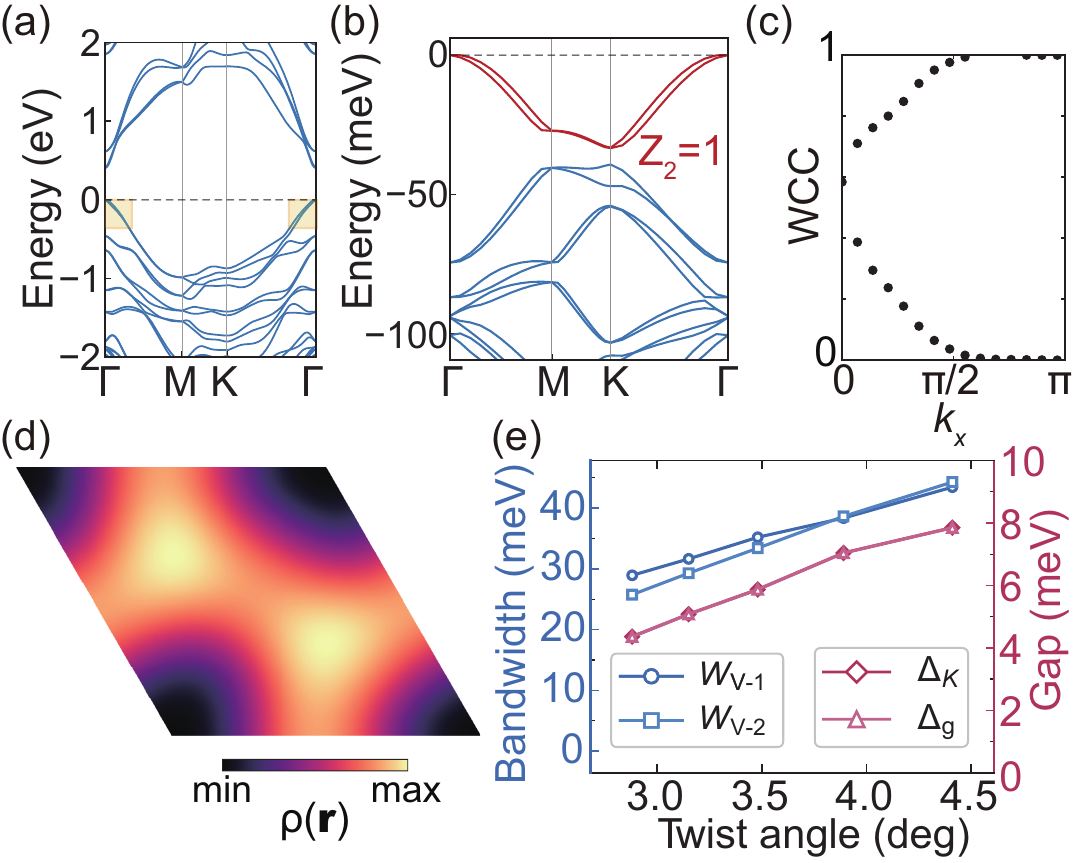}
\caption{
First-principles realization of the proposed moir\'{e} design principle in a representative system, using BiTeBr as an example.
(a) Untwisted bilayer band structure in the energetically favored stacking, with the valence-band maximum at $\Gamma$.
(b) Moir\'{e} bands at $\theta=3.48^\circ$.
(c) Wannier charge centers (WCCs) of the topmost two moir\'{e} valence bands, giving $\mathbb{Z}_2=1$.
(d) Real-space distribution of the topmost four moir\'{e} valence states at the moir\'{e} $K$ point.
(e) Twist-angle dependence of the bandwidths $W_{V-1}$ and $W_{V-2}$ of the two bands adjacent to the Dirac gap, together with the direct $K$-point gap $\Delta_K$ and global gap $\Delta_g$.
Here the topmost moir\'{e} valence band is labeled $V$, lower bands are labeled $V-1$, $V-2$, etc., and $\Delta_g=\min E_{V-1}-\max E_{V-2}$.
}
\label{fig2_BiTeBr}
\end{figure}

We demonstrate this design principle using Janus BiTeBr as an illustrative example. Although a BiTeBr monolayer exhibits strong Rashba splitting due to its intrinsic out-of-plane polarity, the layer-exchanging mirror symmetry $M_z$ of the bilayer configuration strongly suppresses the net layer-asymmetric potential. The Rashba splitting is therefore greatly reduced, restoring the valence-band maximum to the $\Gamma$ point [Fig.~\ref{fig2_BiTeBr}(a)]. These $\Gamma$-valley states are strongly modulated by the stacking-dependent moir\'{e} potential, providing a clean low-energy basis for the geometry-driven honeycomb mechanism.

At $\theta=3.48^\circ$, the reconstructed moir\'{e} potential produces a pair of topmost valence bands well separated from nearby states [Fig.~\ref{fig2_BiTeBr}(b)]. Their Dirac-like dispersion near the moir\'{e} $K$ points reflects the honeycomb connectivity of the A/B moir\'{e} orbitals \cite{SM}, while spin--orbit coupling opens a time-reversal-preserving gap. The Wilson-loop spectrum exhibits partner switching, confirming a nontrivial $\mathbb{Z}_2=1$ invariant [Fig.~\ref{fig2_BiTeBr}(c)]. Real-space wavefunction distributions at the moir\'{e} $K$ point, obtained from the four valence states forming the gapped Dirac manifold, are concentrated in the reconstructed A/B domains [Fig.~\ref{fig2_BiTeBr}(d)], directly visualizing the localized orbitals underlying the honeycomb description~\cite{angeli2021gamma,xu2025twisted}.

Figure~\ref{fig2_BiTeBr}(e) summarizes the twist-angle dependence of the isolated topmost moir\'{e} valence manifold. Reducing $\theta$ enlarges the moir\'{e} period and suppresses the kinetic energy, leading to progressively narrower valence bands. Both the direct valley gap $\Delta_K$ and the global gap $\Delta_g$ decrease toward smaller twist angles, while $\Delta_g$ remains positive, ensuring a well-defined two-band subspace. The evolutions of Wannier charge centres further confirm that the two-band manifold remains topologically nontrivial ($\mathbb{Z}_2=1$) over the full twist-angle range considered. This persistence reflects that changing the twist angle mainly renormalizes the kinetic scale and miniband isolation, while the A/B honeycomb connectivity and symmetry-allowed spin--orbit mass remain intact as long as the gap does not close. Similar symmetry-guided behavior is observed in other triangular-lattice bilayers, including BiTeCl, 1T-CdBr$_2$, ZnI$_2$, and 1H-MoSe$_2$~\cite{SM}, which exhibit the same emergent honeycomb moir\'{e} orbitals and Kane--Mele-type band structures.

\textit{The Kane--Mele diagnosis.---}
To capture the microscopic origin of the moir\'{e} electronic bands, we construct a four-band effective Hamiltonian based on first-principles calculations within a Wannier framework~\cite{marzari2012maximally}. The low-energy model is formulated in an A/B domain basis, reflecting the emergent sublattice structure of the honeycomb moir\'{e} lattice. A key feature of this construction is that the Wannier orbitals are spatially extended over a significant fraction of the moir\'{e} unit cell, leading to strong overlap between neighboring domains. As a result, both hopping amplitudes and spin--orbit couplings are renormalized by the real-space structure of the moir\'{e} states.

Nevertheless, the low-energy sector can be organized as an extended Kane--Mele model,
\begin{equation}
H_{\rm KM}^{\rm ext}=H_{\rm KM}+H_{\rm corr}.
\end{equation}
Here $H_{\rm KM}$ is the standard Kane--Mele Hamiltonian on the emergent honeycomb lattice, while $H_{\rm corr}$ contains additional local spin--orbit terms allowed by the reconstructed moir\'{e} environment. We first write
\begin{align}
H_{\rm KM}
=&\;
t_1\sum_{\langle ij\rangle}c_i^\dagger c_j
+i\lambda_R
\sum_{\langle ij\rangle}
c_i^\dagger
\left[
\left(
\mathbf{s}\times\hat{\mathbf d}_{ij}
\right)_z
\right]
c_j
\nonumber\\
&+
t_2\sum_{\langle\langle ij\rangle\rangle}c_i^\dagger c_j
+i\lambda_{\rm SO}
\sum_{\langle\langle ij\rangle\rangle}
\nu_{ij}c_i^\dagger s_z c_j
+{\rm H.c.}.
\label{eq:H_KM}
\end{align}
The NN hopping $t_1$ generates the leading A--B inter-domain hybridization and gives rise to Dirac crossings at the moir\'{e} $K$ and $K'$ points in the spin-independent limit. The intrinsic NNN spin--orbit coupling $\lambda_{\rm SO}$ provides the Kane--Mele mass that opens a time-reversal-preserving topological gap, while the NN Rashba coupling $\lambda_R$ introduces spin mixing when out-of-plane mirror symmetry is absent.

The correction $H_{\rm corr}$ describes additional same-sublattice NNN spin--orbit channels with in-plane spin components. Such terms are absent in the mirror-symmetric Kane--Mele limit, but are allowed in the reconstructed moir\'{e} structure where $M_z$ is no longer an exact symmetry. We parameterize these $M_z$-breaking channels by $\lambda_{\rm rad}^{\rm NNN}$ and $\lambda_{\rm in}^{\rm NNN}$, with the explicit bond-dependent form given in the Supplemental Material~\cite{SM}.

\begin{figure}[htbp]
\includegraphics[width=0.98\linewidth]{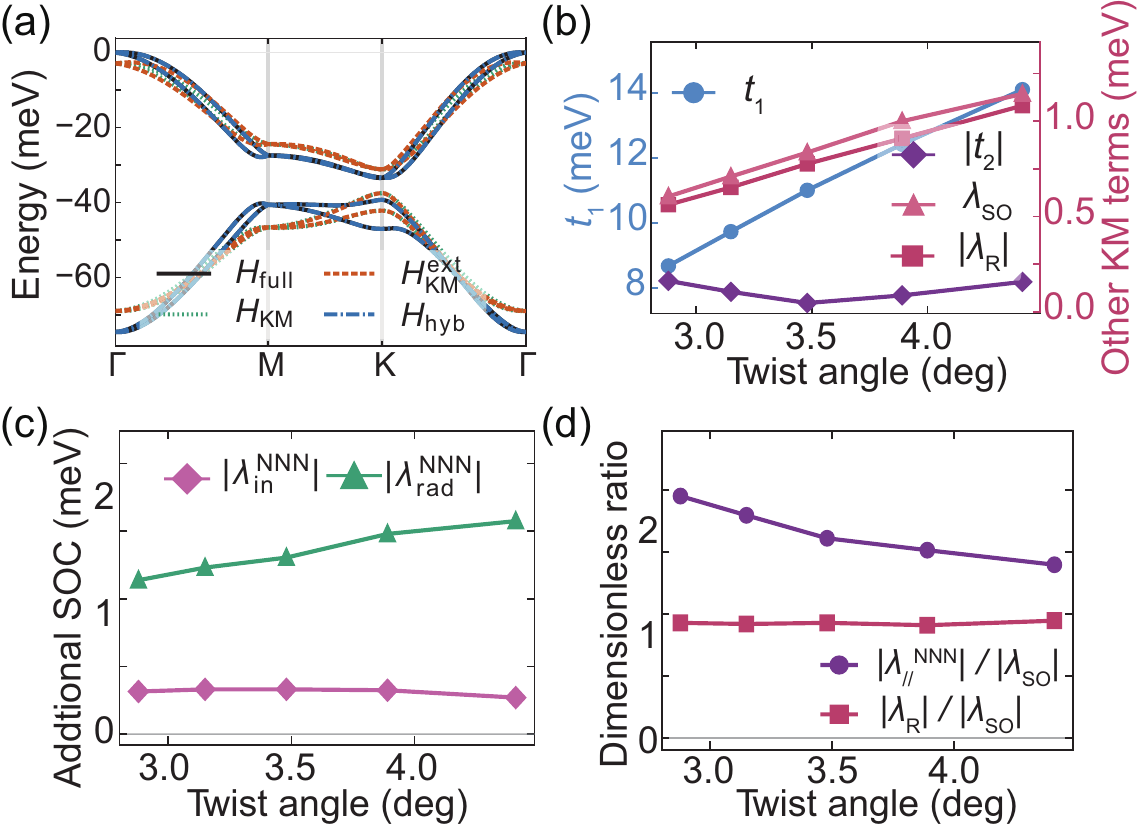}
    \caption{
    Extended Kane--Mele diagnosis of the moir\'{e} bands.
    (a) Band comparison for the BiTeBr moir\'{e} superlattice at $\theta=3.48^\circ$ between the first-principles band structure and a hierarchy of effective models, including the standard Kane--Mele model $H_{\rm KM}$, the extended model $H_{\rm KM}^{\rm ext}$, and the hybrid model $H_{\rm hyb}=H_{\rm KM}^{\rm ext}+H_{\rm far}$. Here $H_{\rm far}$ restores farther-neighbor hopping processes.
    (b) Twist-angle evolution of the standard Kane-Mele model parameters: $t_1$, $|t_2|$, $\lambda_{\rm SO}$, and $|\lambda_R|$.
    (c) Same-sublattice NNN in-plane spin-orbit channels $|\lambda_{\rm rad}^{\rm NNN}|$ and $|\lambda_{\rm in}^{\rm NNN}|$.
    (d) Ratios $|\lambda_{\parallel}^{\rm NNN}|/|\lambda_{\rm SO}|$ and $|\lambda_R|/|\lambda_{\rm SO}|$, with $|\lambda_{\parallel}^{\rm NNN}|=\sqrt{(\lambda_{\rm rad}^{\rm NNN})^2+(\lambda_{\rm in}^{\rm NNN})^2}$, showing sizable corrections beyond the standard Kane--Mele model.
    }
    \label{fig:angle_evolution}
\end{figure}

Figure~\ref{fig:angle_evolution}(a) compares the first-principles band structure at $\theta=3.48^\circ$ with a hierarchy of effective models. The minimal Kane--Mele model captures the Dirac gap structure and the topology of the isolated manifold. The extended model $H_{\rm KM}^{\rm ext}=H_{\rm KM}+H_{\rm corr}$ incorporates additional local symmetry-allowed spin--orbit channels, while the hybrid model $H_{\rm hyb}=H_{\rm KM}^{\rm ext}+H_{\rm far}$ restores farther-neighbor hopping processes and quantitatively reproduces the full band dispersion.

The extracted parameters vary smoothly with twist angle [Fig.~\ref{fig:angle_evolution}(b)], revealing a robust separation of energy scales. The nearest-neighbor hopping $t_1$ dominates as the kinetic energy scale of the emergent honeycomb lattice and increases with increasing twist angle, reflecting enhanced inter-domain hybridization at shorter moir\'{e} periods. In contrast, $|t_2|$, $\lambda_{\rm SO}$, and $\lambda_R$ remain at the meV scale throughout the studied angle range.

Importantly, the intrinsic spin--orbit coupling $\lambda_{\rm SO}$ remains the dominant source of the topological mass at zero field, ensuring a robust $\mathbb{Z}_2=1$ phase across the twist-angle range considered. The Rashba coupling $\lambda_R$ competes with $\lambda_{\rm SO}$ by inducing spin mixing, but does not close the bulk gap in the zero-field structures, so that the system remains in the quantum spin Hall phase.

In Janus systems, additional NNN in-plane spin--orbit channels become sizable [Fig.~\ref{fig:angle_evolution}(c)], providing quantitative corrections to the spin texture and band dispersion. By contrast, centrosymmetric systems such as 1T-phase CdBr$_2$ and ZnI$_2$ exhibit much weaker spin-dependent splittings and lie closer to the minimal Kane--Mele limit. These representative systems therefore span distinct spin--orbit hierarchies, from Janus compounds with sizable Rashba and in-plane spin--orbit channels to centrosymmetric bilayers in which the minimal Kane--Mele description is nearly recovered \cite{SM}. Across the cases where the topmost two-band manifold is well isolated, the $\mathbb{Z}_2$ invariant remains nontrivial, indicating that the extended Kane--Mele framework describes a robust topological regime across different material classes, as summarized in Table~\ref{tab:KM_characteristic_scales}.

\begin{table}[htbp]
\centering
\caption{
Kane--Mele parameters at $\theta=3.48^\circ$ for additional representative triangular-lattice bilayers, in meV. The last column gives the $\mathbb{Z}_2$ invariant of the topmost two moir\'{e} valence bands when isolated; ``--'' indicates that the corresponding two-band manifold is not well isolated.
}
\label{tab:KM_characteristic_scales}
\renewcommand{\arraystretch}{1.1}
\begin{ruledtabular}
\begin{tabular}{lccccc}
Material & $t_1$ & $t_2$ & $\lambda_{\rm SO}$ & $\lambda_R$ & $\mathbb{Z}_2$ \\
\hline
ZnI$_2$  &  7.850 &  1.062 & 0.206 & -0.003 & 1 \\
CdBr$_2$ &  2.980 &  0.032 & 0.054 & -0.013 & 1 \\
MoSe$_2$ &  3.376 & -0.358 & 0.003 &  0.009 & -- \\
BiTeCl   & 10.243 & -0.197 & 0.848 &  1.221 & 1 \\
\end{tabular}
\end{ruledtabular}
\end{table}

\begin{figure}[htbp]
\includegraphics[width=0.98\linewidth]{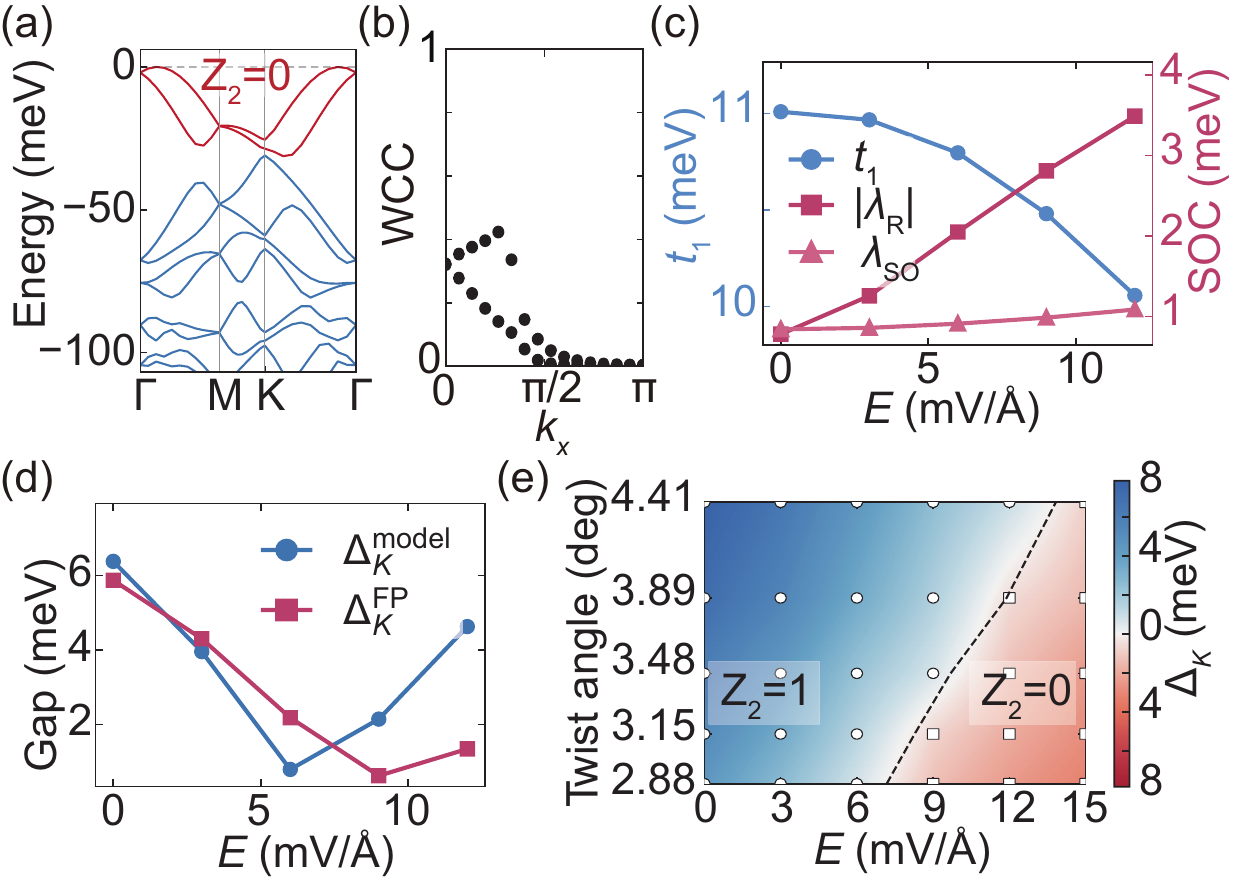}
    \caption{
    Electric-field-driven topological transition in twisted BiTeBr bilayers.
    (a) Moir\'{e} bands and (b) WCC evolution for the topmost two moir\'{e} valence bands at $\theta=3.48^\circ$ and $E=15$~mV/${\rm \AA}$.
    (c) Electric-field dependence of representative Kane--Mele parameters: $t_1$, $\lambda_{\rm SO}$, and $\lambda_R$.
    (d) Comparison between the extended Kane--Mele estimate and first-principles result for the $K$-point gap.
    (e) Twist-angle--electric-field phase diagram. Open circles and squares denote calculated $\mathbb{Z}_2=1$ and $\mathbb{Z}_2=0$ points, respectively; the dashed line marks the estimated gap-closing boundary.
    }
    \label{fig:efield}
\end{figure}

\textit{Electric-field-tuned topological transition.---}
The spatially extended moir\'{e} states are naturally susceptible to external electric fields, providing a convenient knob to manipulate their topological properties. At a representative twist angle $\theta=3.48^\circ$, increasing an out-of-plane electric field drives the moir\'{e} $K$-point gap to close and reopen. At $E=15$ mV/\AA, the reopened gap corresponds to a trivial phase. The corresponding moir\'{e} bands are shown in Fig.~\ref{fig:efield}(a), while Wannier charge centers confirm a change of the topological invariant from $\mathbb{Z}_2=1$ to $\mathbb{Z}_2=0$ [Fig.~\ref{fig:efield}(b)], consistent with the field-dependent band evolution~\cite{SM}.

Analysis of the extended Kane--Mele model parameters reveals that the applied electric field primarily enhances the NN Rashba coupling $\lambda_R$, while leaving the NN hopping $t_1$ and intrinsic Kane--Mele coupling $\lambda_{\rm SO}$ nearly unchanged [Fig.~\ref{fig:efield}(c)]. This separation of responses reflects their distinct microscopic origins. The weak field dependence of $\lambda_{\rm SO}$ is consistent with its dominant atomic-orbital character, whereas $\lambda_R$ is directly tied to field-induced inversion asymmetry in the moir\'{e} environment. The electric field therefore does not primarily destroy the honeycomb connectivity; instead, it tunes the spin--orbit mass competition within the same low-energy manifold. The $K$-point gap obtained from the extended Kane--Mele model reproduces the trend of the first-principles result and provides an estimate of the critical field [Fig.~\ref{fig:efield}(d)].

Extending this analysis across multiple twist angles, we construct the twist-angle--electric-field phase diagram [Fig.~\ref{fig:efield}(e)]. Smaller twist angles require lower critical fields for the topological transition, reflecting the enhanced field sensitivity of more extended moir\'{e} states. The gap ${\Delta}_K$ illustrates the evolution from the zero-field-connected $\mathbb{Z}_2=1$ quantum spin Hall phase to the field-driven trivial $\mathbb{Z}_2=0$ phase, with the dashed line indicating the estimated gap-closing boundary. These results demonstrate that the electric-field-driven gap closing and reopening are governed by the competition between the intrinsic Kane--Mele coupling $\lambda_{\rm SO}$ and the field-enhanced Rashba coupling $\lambda_R$.

\textit{Discussion.---}
The coexistence of nontrivial $\mathbb{Z}_2$ topology and narrow moir\'{e} bandwidths places these systems in a correlated topological regime, where interactions act within the same emergent honeycomb manifold that supports the Kane--Mele mass and may renormalize the quantum spin Hall phase, stabilize symmetry-broken descendants, or promote fractionalized states~\cite{meng2010quantum, hohenadler2012quantum, rachel2018interacting}. More broadly, the same geometry--symmetry principle may extend to magnetic triangular bilayers hosting Haldane-type Chern bands, or to other stacking landscapes with distinct emergent lattice geometries~\cite{haldane1988model, wang2023quantum}.

\textit{Summary.---}
We identify a geometry--symmetry route to $\mathbb{Z}_2$ topology in twisted triangular-lattice bilayers, where symmetry-related stacking minima confine $\Gamma$-valley band-edge states into A/B moir\'{e} orbitals forming an emergent honeycomb lattice. First-principles calculations reveal a robust quantum spin Hall phase over a broad twist-angle range, described by an extended Kane--Mele model and tunable through an electric-field-driven topological transition. Its occurrence across representative triangular-lattice moir\'{e} systems establishes a broadly applicable framework for tunable moir\'{e} quantum spin Hall materials.

\paragraph{Acknowledgments.} This work was supported by the National Key R\&D Program of China (Nos. 2024YFA1408400, 2023YFA1607400, 2022YFA1403800), the National Natural Science Foundation of China (Grant Nos. 12274436, 11925408, and 11921004), the Science Center of the National Natural Science Foundation of China (Grant No. 12188101), and the Beijing Municipal Science \& Technology Commission, Administrative Commission of Zhongguancun Science Park No. Z251100003625025. H. Weng acknowledges support from the New Cornerstone Science Foundation (XPLORER PRIZE). J. Li acknowledges support from the China National Postdoctoral Program for Innovative Talents (Grant No. BX20220334). 

\textit{Note added.---} After completion of this work, we became aware of a related independent work~\cite{pi2026engineering}.

\bibliography{reference.bib}

@article{zhang2024universal,
  title={Universal Moir{\'e}-Model-Building Method without Fitting: Application to Twisted {MoTe$_2$} and {WSe$_2$}},
  author={Zhang, Yan and Pi, Hanqi and Liu, Jiaxuan and Miao, Wangqian and Qi, Ziyue and Regnault, Nicolas and Weng, Hongming and Dai, Xi and Bernevig, B Andrei and Wu, Quansheng and others},
  journal={arXiv preprint arXiv:2411.08108},
  url={https://doi.org/10.48550/arXiv.2411.08108},
  year={2024}
}

@article{cai2023signatures,
  title={Signatures of fractional quantum anomalous {Hall} states in twisted {MoTe$_2$}},
  author={Cai, Jiaqi and Anderson, Eric and Wang, Chong and Zhang, Xiaowei and Liu, Xiaoyu and Holtzmann, William and Zhang, Yinong and Fan, Fengren and Taniguchi, Takashi and Watanabe, Kenji and others},
  journal={Nature},
  volume={622},
  number={7981},
  pages={63--68},
  year={2023},
  publisher={Nature Publishing Group UK London}
}

@article{bistritzer2011moire,
author = {Rafi Bistritzer  and Allan H. MacDonald },
title = {Moir{\'e} bands in twisted double-layer graphene},
journal = {Proc. Natl. Acad. Sci.},
volume = {108},
number = {30},
pages = {12233-12237},
year = {2011},
doi = {10.1073/pnas.1108174108},
URL = {https://www.pnas.org/doi/abs/10.1073/pnas.1108174108},
}

@article{cao2020tunable,
  title={Tunable correlated states and spin-polarized phases in twisted bilayer--bilayer graphene},
  author={Cao, Yuan and Rodan-Legrain, Daniel and Rubies-Bigorda, Oriol and Park, Jeong Min and Watanabe, Kenji and Taniguchi, Takashi and Jarillo-Herrero, Pablo},
  journal={Nature},
  volume={583},
  number={7815},
  pages={215--220},
  year={2020},
  url={https://www.nature.com/articles/s41586-020-2260-6},
  publisher={Nature Publishing Group UK London}
}

@article{cao2018unconventional,
  title={Unconventional superconductivity in magic-angle graphene superlattices},
  author={Cao, Yuan and Fatemi, Valla and Fang, Shiang and Watanabe, Kenji and Taniguchi, Takashi and Kaxiras, Efthimios and Jarillo-Herrero, Pablo},
  journal={Nature},
  volume={556},
  number={7699},
  pages={43--50},
  year={2018},
  publisher={Nature Publishing Group UK London},
  url={https://www.nature.com/articles/nature26160}
}

@article{zhang2024polarization,
  title={Polarization-driven band topology evolution in twisted {MoTe$_2$} and {WSe$_2$}},
  author={Zhang, Xiao-Wei and Wang, Chong and Liu, Xiaoyu and Fan, Yueyao and Cao, Ting and Xiao, Di},
  journal={Nat. Commun.},
  volume={15},
  number={1},
  pages={4223},
  year={2024},
  doi={https://www.nature.com/articles/s41467-024-48511-x},
  publisher={Nature Publishing Group UK London}
}

@article{wang2024fractional,
  title={Fractional {Chern} insulator in twisted bilayer {MoTe$_2$}},
  author={Wang, Chong and Zhang, Xiao-Wei and Liu, Xiaoyu and He, Yuchi and Xu, Xiaodong and Ran, Ying and Cao, Ting and Xiao, Di},
  journal={Phys. Rev. Lett.},
  volume={132},
  number={3},
  pages={036501},
  year={2024},
  url={https://journals.aps.org/prl/abstract/10.1103/PhysRevLett.132.036501},
  publisher={APS}
}

@article{redekop2024direct,
  title={Direct magnetic imaging of fractional {Chern} insulators in twisted {MoTe$_2$}},
  author={Redekop, Evgeny and Zhang, Canxun and Park, Heonjoon and Cai, Jiaqi and Anderson, Eric and Sheekey, Owen and Arp, Trevor and Babikyan, Grigory and Salters, Samuel and Watanabe, Kenji and others},
  journal={Nature},
  volume={635},
  number={8039},
  pages={584--589},
  year={2024},
  url={https://www.nature.com/articles/s41586-024-08153-x},
  publisher={Nature Publishing Group UK London}
}

@article{xu2023observation,
  title={Observation of integer and fractional quantum anomalous {Hall} effects in twisted bilayer {MoTe$_2$}},
  author={Xu, Fan and Sun, Zheng and Jia, Tongtong and Liu, Chang and Xu, Cheng and Li, Chushan and Gu, Yu and Watanabe, Kenji and Taniguchi, Takashi and Tong, Bingbing and others},
  journal={Phys. Rev. X},
  volume={13},
  number={3},
  pages={031037},
  year={2023},
  url={https://journals.aps.org/prx/abstract/10.1103/PhysRevX.13.031037},
  publisher={APS}
}

@article{xu2025multiple,
  title={Multiple {Chern} bands in twisted {MoTe$_2$} and possible {non-Abelian} states},
  author={Xu, Cheng and Mao, Ning and Zeng, Tiansheng and Zhang, Yang},
  journal={Phys. Rev. Lett.},
  volume={134},
  number={6},
  pages={066601},
  year={2025},
  url={https://journals.aps.org/prl/abstract/10.1103/PhysRevLett.134.066601},
  publisher={APS}
}

@article{jia2024moire,
  title={Moir{\'e} fractional {Chern} insulators. {I}. {First}-principles calculations and continuum models of twisted bilayer {MoTe$_2$}},
  author={Jia, Yujin and Yu, Jiabin and Liu, Jiaxuan and Herzog-Arbeitman, Jonah and Qi, Ziyue and Pi, Hanqi and Regnault, Nicolas and Weng, Hongming and Bernevig, B Andrei and Wu, Quansheng},
  journal={Phys. Rev. B},
  volume={109},
  number={20},
  pages={205121},
  year={2024},
  url={https://journals.aps.org/prb/abstract/10.1103/PhysRevB.109.205121},
  publisher={APS}
}

@article{zeng2023thermodynamic,
  title={Thermodynamic evidence of fractional {Chern} insulator in moir{\'e} {MoTe$_2$}},
  author={Zeng, Yihang and Xia, Zhengchao and Kang, Kaifei and Zhu, Jiacheng and Kn{\"u}ppel, Patrick and Vaswani, Chirag and Watanabe, Kenji and Taniguchi, Takashi and Mak, Kin Fai and Shan, Jie},
  journal={Nature},
  volume={622},
  number={7981},
  pages={69--73},
  year={2023},
  url={https://www.nature.com/articles/s41586-023-06452-3},
  publisher={Nature Publishing Group UK London}
}

@article{xia2025superconductivity,
  title={Superconductivity in twisted bilayer {WSe$_2$}},
  author={Xia, Yiyu and Han, Zhongdong and Watanabe, Kenji and Taniguchi, Takashi and Shan, Jie and Mak, Kin Fai},
  journal={Nature},
  volume={637},
  number={8047},
  pages={833--838},
  year={2025},
  url={https://www.nature.com/articles/s41586-024-08116-2},
  publisher={Nature Publishing Group UK London}
}

@article{guo2025superconductivity,
  title={Superconductivity in 5.0° twisted bilayer {WSe$_2$}},
  author={Guo, Yinjie and Pack, Jordan and Swann, Joshua and Holtzman, Luke and Cothrine, Matthew and Watanabe, Kenji and Taniguchi, Takashi and Mandrus, David G and Barmak, Katayun and Hone, James and others},
  journal={Nature},
  volume={637},
  number={8047},
  pages={839--845},
  year={2025},
  url={https://www.nature.com/articles/s41586-024-08381-1},
  publisher={Nature Publishing Group UK London}
}

@article{carr2020electronic,
  title={Electronic-structure methods for twisted moir{\'e} layers},
  author={Carr, Stephen and Fang, Shiang and Kaxiras, Efthimios},
  journal={Nat. Rev. Mater.},
  volume={5},
  number={10},
  pages={748--763},
  year={2020},
  url={https://www.nature.com/articles/s41578-020-0214-0},
  publisher={Nature Publishing Group UK London}
}

@article{kennes2021moire,
  title={Moir{\'e} heterostructures as a condensed-matter quantum simulator},
  author={Kennes, Dante M and Claassen, Martin and Xian, Lede and Georges, Antoine and Millis, Andrew J and Hone, James and Dean, Cory R and Basov, DN and Pasupathy, Abhay N and Rubio, Angel},
  journal={Nat. Phys.},
  volume={17},
  number={2},
  pages={155--163},
  year={2021},
  url={https://www.nature.com/articles/s41567-020-01154-3},
  publisher={Nature Publishing Group UK London}
}

@article{carr2017twistronics,
  title={Twistronics: Manipulating the electronic properties of two-dimensional layered structures through their twist angle},
  author={Carr, Stephen and Massatt, Daniel and Fang, Shiang and Cazeaux, Paul and Luskin, Mitchell and Kaxiras, Efthimios},
  journal={Phys. Rev. B},
  volume={95},
  number={7},
  pages={075420},
  year={2017},
  url={https://journals.aps.org/prb/abstract/10.1103/PhysRevB.95.075420},
  publisher={APS}
}

@article{andrei2021marvels,
  title={The marvels of moir{\'e} materials},
  author={Andrei, Eva Y and Efetov, Dmitri K and Jarillo-Herrero, Pablo and MacDonald, Allan H and Mak, Kin Fai and Senthil, T and Tutuc, Emanuel and Yazdani, Ali and Young, Andrea F},
  journal={Nat. Rev. Mater.},
  volume={6},
  number={3},
  pages={201--206},
  year={2021},
  url={https://www.nature.com/articles/s41578-021-00284-1},
  publisher={Nature Publishing Group UK London}
}

@article{bradlyn2017topological,
  title={Topological quantum chemistry},
  author={Bradlyn, Barry and Elcoro, Luis and Cano, Jennifer and Vergniory, Maia G and Wang, Zhijun and Felser, Claudia and Aroyo, Mois I and Bernevig, B Andrei},
  journal={Nature},
  volume={547},
  number={7663},
  pages={298--305},
  year={2017},
  url={https://www.nature.com/articles/nature23268},
  publisher={Nature Publishing Group UK London}
}

@article{yang2024topological,
  title={Topological minibands and interaction driven quantum anomalous {Hall} state in topological insulator based moir{\'e} heterostructures},
  author={Yang, Kaijie and Xu, Zian and Feng, Yanjie and Schindler, Frank and Xu, Yuanfeng and Bi, Zhen and Bernevig, B Andrei and Tang, Peizhe and Liu, Chao-Xing},
  journal={Nat. Commun.},
  volume={15},
  number={1},
  pages={2670},
  year={2024},
  publisher={Nature Publishing Group UK London}
}

@article{stern2016fractional,
  title={Fractional topological insulators: a pedagogical review},
  author={Stern, Ady},
  journal={Annu. Rev. Condens. Matter Phys},
  volume={7},
  number={1},
  pages={349--368},
  year={2016},
  publisher={Annual Reviews}
}

@article{kane2005quantum,
  title={Quantum spin {Hall} effect in graphene},
  author={Kane, Charles L and Mele, Eugene J},
  journal={Phys. Rev. Lett.},
  volume={95},
  number={22},
  pages={226801},
  year={2005},
  publisher={APS}
}

@article{song2018quantitative,
  title={Quantitative mappings between symmetry and topology in solids},
  author={Song, Zhida and Zhang, Tiantian and Fang, Zhong and Fang, Chen},
  journal={Nat. Commun.},
  volume={9},
  number={1},
  pages={3530},
  year={2018},
  publisher={Nature Publishing Group UK London}
}

@article{liu2025symmetry,
  title={Symmetry-enforced Moir\'e Topology},
  author={Liu, Yunzhe and Angerhofer, Ethan and Yang, Kaijie and Liu, Chao-Xing and Yu, Jiabin},
  journal={arXiv preprint arXiv:2509.06906},
  year={2025}
}

@article{meng2010quantum,
  title={Quantum spin liquid emerging in two-dimensional correlated {Dirac} fermions},
  author={Meng, ZY and Lang, TC and Wessel, S and Assaad, FF and Muramatsu, A},
  journal={Nature},
  volume={464},
  number={7290},
  pages={847--851},
  year={2010},
  publisher={Nature Publishing Group UK London}
}

@article{hohenadler2012quantum,
  title={Quantum phase transitions in the {Kane-Mele-Hubbard} model},
  author={Hohenadler, Martin and Meng, ZY and Lang, TC and Wessel, S and Muramatsu, A and Assaad, FF},
  journal={Phys. Rev. B},
  volume={85},
  number={11},
  pages={115132},
  year={2012},
  publisher={APS}
}

@article{zhang2009topological,
  title={Topological insulators in {Bi$_2$Se$_3$}, {Bi$_2$Te$_3$} and {Sb$_2$Te$_3$} with a single {Dirac} cone on the surface},
  author={Zhang, Haijun and Liu, Chao-Xing and Qi, Xiao-Liang and Dai, Xi and Fang, Zhong and Zhang, Shou-Cheng},
  journal={Nat. Phys.},
  volume={5},
  number={6},
  pages={438--442},
  year={2009},
  publisher={Nature Publishing Group UK London}
}

@article{bernevig2006quantum,
  title={Quantum spin {Hall} effect and topological phase transition in {HgTe} quantum wells},
  author={Bernevig, B Andrei and Hughes, Taylor L and Zhang, Shou-Cheng},
  journal={science},
  volume={314},
  number={5806},
  pages={1757--1761},
  year={2006},
  publisher={American Association for the Advancement of Science}
}

@article{galanakis2002origin,
  title={Origin and properties of the gap in the half-ferromagnetic {Heusler} alloys},
  author={Galanakis, I and Dederichs, PH and Papanikolaou, NJPRB},
  journal={Phys. Rev. B},
  volume={66},
  number={13},
  pages={134428},
  year={2002},
  publisher={APS}
}

@article{bartel2019new,
  title={New tolerance factor to predict the stability of perovskite oxides and halides},
  author={Bartel, Christopher J and Sutton, Christopher and Goldsmith, Bryan R and Ouyang, Runhai and Musgrave, Charles B and Ghiringhelli, Luca M and Scheffler, Matthias},
  journal={Sci. Adv.},
  volume={5},
  number={2},
  pages={eaav0693},
  year={2019},
  publisher={American Association for the Advancement of Science}
}

@article{tang2021geometric,
  title={Geometric origins of topological insulation in twisted layered semiconductors},
  author={Tang, Hao and Carr, Stephen and Kaxiras, Efthimios},
  journal={Phys. Rev. B},
  volume={104},
  number={15},
  pages={155415},
  year={2021},
  publisher={APS}
}

@article{rachel2018interacting,
  title={Interacting topological insulators: a review},
  author={Rachel, Stephan},
  journal={Rep. Prog. Phys.},
  volume={81},
  number={11},
  pages={116501},
  year={2018},
  publisher={IOP Publishing}
}

@article{zhang2020flat,
  title={Flat bands in twisted bilayer transition metal dichalcogenides},
  author={Zhang, Zhiming and Wang, Yimeng and Watanabe, Kenji and Taniguchi, Takashi and Ueno, Keiji and Tutuc, Emanuel and LeRoy, Brian J},
  journal={Nat. Phys.},
  volume={16},
  number={11},
  pages={1093--1096},
  year={2020},
  publisher={Nature Publishing Group UK London}
}

@article{kang2024evidence,
  title={Evidence of the fractional quantum spin Hall effect in moir{\'e} {MoTe$_2$}},
  author={Kang, Kaifei and Shen, Bowen and Qiu, Yichen and Zeng, Yihang and Xia, Zhengchao and Watanabe, Kenji and Taniguchi, Takashi and Shan, Jie and Mak, Kin Fai},
  journal={Nature},
  volume={628},
  number={8008},
  pages={522--526},
  year={2024},
  publisher={Nature Publishing Group UK London}
}

@article{haldane1988model,
  title={Model for a quantum {Hall} effect without {Landau} levels: Condensed-matter realization of the ``parity anomaly''},
  author={Haldane, F Duncan M},
  journal={Phys. Rev. Lett.},
  volume={61},
  number={18},
  pages={2015},
  year={1988},
  publisher={APS}
}

@article{wang2023quantum,
  title={Quantum states and intertwining phases in kagome materials},
  author={Wang, Yaojia and Wu, Heng and McCandless, Gregory T and Chan, Julia Y and Ali, Mazhar N},
  journal={Nat. Rev. Phys.},
  volume={5},
  number={11},
  pages={635--658},
  year={2023},
  publisher={Nature Publishing Group UK London}
}

@article{angeli2021gamma,
  title={{$\Gamma$} valley transition metal dichalcogenide moir{\'e} bands},
  author={Angeli, Mattia and MacDonald, Allan H},
  journal={Proc. Natl. Acad. Sci. },
  volume={118},
  number={10},
  pages={e2021826118},
  year={2021},
  publisher={National Academy of Sciences}
}

@article{wu2019topological,
  title={Topological insulators in twisted transition metal dichalcogenide homobilayers},
  author={Wu, Fengcheng and Lovorn, Timothy and Tutuc, Emanuel and Martin, Ivar and MacDonald, AH},
  journal={Phys. Rev. Lett.},
  volume={122},
  number={8},
  pages={086402},
  year={2019},
  publisher={APS}
}

@article{wang2020correlated,
  title={Correlated electronic phases in twisted bilayer transition metal dichalcogenides},
  author={Wang, Lei and Shih, En-Min and Ghiotto, Augusto and Xian, Lede and Rhodes, Daniel A and Tan, Cheng and Claassen, Martin and Kennes, Dante M and Bai, Yusong and Kim, Bumho and others},
  journal={Nat. Mater.},
  volume={19},
  number={8},
  pages={861--866},
  year={2020},
  publisher={Nature Publishing Group UK London}
}

@article{xu2025twisted,
author = {Xu, Xilong and Wang, Haonan and Yang, Li},
title = {Twisted Type-{II} {Rashba} Homobilayers: A Platform for Tunable Topological Moiré Flat Bands},
journal = {Adv. Funct. Mater.},
volume = {35},
number = {40},
pages = {2425454},
doi = {https://doi.org/10.1002/adfm.202425454},
url = {https://advanced.onlinelibrary.wiley.com/doi/abs/10.1002/adfm.202425454},
year = {2025}
}

@article{pi2026engineering,
  title={Engineering topological flat bands in {$\Gamma$}-valley moir\'e systems with {Ising-type SOC}: twisted {1T-ZrS$_2 $ and 1T-SnSe$_2$}},
  author={Pi, Hanqi and Kwan, Yves H and Hu, Haoyu and Jiang, Yi and C{\u{a}}lug{\u{a}}ru, Dumitru and Shan, Jie and Mak, Kin Fai and Ugeda, Miguel M and Efetov, Dmitri K and Vergniory, Maia G and others},
  journal={arXiv:2605.13984},
  year={2026}
}

@article{kruthoff2017topological,
  title = {Topological Classification of Crystalline Insulators through Band Structure Combinatorics},
  author = {Kruthoff, Jorrit and de Boer, Jan and van Wezel, Jasper and Kane, Charles L. and Slager, Robert-Jan},
  journal = {Phys. Rev. X},
  volume = {7},
  issue = {4},
  pages = {041069},
  numpages = {23},
  year = {2017},
  month = {Dec},
  publisher = {American Physical Society},
  doi = {10.1103/PhysRevX.7.041069},
  url = {https://link.aps.org/doi/10.1103/PhysRevX.7.041069}
}

@article{marzari2012maximally,
  title = {Maximally localized {Wannier} functions: Theory and applications},
  author = {Marzari, Nicola and Mostofi, Arash A. and Yates, Jonathan R. and Souza, Ivo and Vanderbilt, David},
  journal = {Rev. Mod. Phys.},
  volume = {84},
  issue = {4},
  pages = {1419--1475},
  numpages = {0},
  year = {2012},
  month = {Oct},
  publisher = {American Physical Society},
  doi = {10.1103/RevModPhys.84.1419},
  url = {https://link.aps.org/doi/10.1103/RevModPhys.84.1419}
}

@misc{SM,
	note={
        Supplementary Materials include computational methods, additional electronic band structures, model Hamiltonian analyses, and the topological properties of representative moir\'e two-dimensional materials. Electric-field-induced phase transitions are discussed for BiTeBr.
    }
}
\end{document}